\documentclass{article}

\usepackage[nonatbib, preprint]{cig}

\usepackage[utf8]{inputenc}
\usepackage[T1]{fontenc}

\usepackage{amsmath, amssymb, amsfonts, mathtools, bm}
\usepackage{nicefrac}
\usepackage{bbm}
\usepackage{siunitx}
\usepackage{pifont}
\usepackage{algorithm}
\usepackage{algpseudocode}
\usepackage{amsthm}
\usepackage{calc}
\usepackage{cite}
\usepackage{xcolor}
\definecolor{darkblue}{rgb}{0, 0, 0.85}
\definecolor{lightgreen}{rgb}{.85,1,.85}
\definecolor{lightred}{rgb}{1,.85,.85}
\definecolor{lightblue}{rgb}{.85,.85,1}
\definecolor{pink}{HTML}{EB346F}
\definecolor{cvprblue}{rgb}{0.21,0.49,0.74}

\usepackage{hyperref}
\hypersetup{
    colorlinks = true,
    citecolor = blue,
    linkcolor = black
}

\usepackage{graphicx}
\usepackage{tabularx, threeparttable, tabularray}
\UseTblrLibrary{booktabs}
\usepackage{booktabs}
\usepackage{multirow}
\usepackage{wrapfig}
\usepackage{arydshln}
\usepackage{hhline}
\usepackage{caption}
\usepackage{diagbox}

\usepackage{microtype}
\usepackage{setspace}
\usepackage{lipsum}
\usepackage{soul}

\usepackage{tikz}

\theoremstyle{plain}

\newcommand{\hlgreen}[1]{{\sethlcolor{lightgreen}\hl{#1}}}

\newcommand{\hlblue}[1]{{\sethlcolor{lightblue}\hl{#1}}}

\DeclarePairedDelimiterX{\infdivx}[2]{(}{)}{#1\;\delimsize\|\;#2}

\renewcommand{\vec}[1]{\bm{#1}}

\newcommand{\X}{{\vec{X}}}  
  
\newcommand{\Z}{{\vec{Z}}}  
\newcommand{\Q}{{\vec{Q}}}

\def\x{\vec{x}}

\def\s{\vec{s}}  
\def\n{\vec{n}}

\def\P{\mathbb{P}}
\def\Q{\mathbb{Q}}

\usepackage{pifont}

\title{Deep Parameter Interpolation for Scalar Conditioning}

\vspace{5em}
\author{%
\normalsize Chicago~Y.~Park$^1$ \quad Michael~T.~McCann$^2$ \quad Cristina~Garcia-Cardona$^2$ \\[0.7em] Brendt~Wohlberg$^2$ \quad Ulugbek~S.~Kamilov$^3$\\[0.7em]
\small $^1$\textnormal{WashU} \quad $^2$\textnormal{Los Alamos National Laboratory} \quad $^3$\textnormal{UW--Madison}\\ [0.7em]
\footnotesize \texttt{chicago@wustl.edu}\\[0.2em]
\footnotesize \texttt{\{mccann,  cgarciac, brendt\}@lanl.gov}\\[0.2em]
\footnotesize \texttt{kamilov@wisc.edu}
}

\begin{document}

\maketitle

\begin{abstract}
    We propose deep parameter interpolation (DPI), a general-purpose method for transforming an existing deep neural network architecture into one that accepts an additional scalar input. Recent deep generative models, including diffusion models and flow matching, employ a single neural network to learn a time- or noise level-dependent vector field. Designing a network architecture to accurately represent this vector field is challenging because the network must integrate information from two different sources: a high-dimensional vector (usually an image) and a scalar. Common approaches either encode the scalar as an additional image input or combine scalar and vector information in specific network components, which restricts architecture choices. Instead, we propose to maintain two learnable parameter sets within a single network and to introduce the scalar dependency by dynamically interpolating between the parameter sets based on the scalar value during training and sampling. DPI is a simple, architecture-agnostic method for adding scalar dependence to a neural network. We demonstrate that our method improves denoising performance and enhances sample quality for both diffusion and flow matching models, while achieving computational efficiency comparable to standard scalar conditioning techniques. Code is available at \href{https://github.com/wustl-cig/parameter_interpolation}{\textcolor{blue}{\texttt{https://github.com/wustl-cig/parameter\_interpolation}}}.
\end{abstract}

\section{Introduction}
\label{sec:intro}

Generative modeling has made significant progress through the development of diffusion models~\cite{ho_NEURIPS2020_ddpm, song2021sde} and flow matching methods~\cite{lipman2022flowmatching1, albergo2023flowmatching2, liu2022flowmatching3}. Both frameworks generate complex data by progressively transforming samples from tractable source distributions --- such as Gaussian noise --- into structured samples. 
Diffusion models achieve this through iterative denoising guided by learned score functions, typically formulated as stochastic differential equations (SDEs). In contrast, flow matching models learn deterministic dynamics along probability paths, often formulated as ordinary differential equations (ODEs), to continuously transport samples from a source distribution toward the data distribution.

To model this progressive transformation, diffusion and flow matching frameworks typically employ a single neural network that operates across all sampling steps~\cite{song2019generative, ho_NEURIPS2020_ddpm, song2021sde, lipman2022flowmatching1, park2024randomwalks}.
This network is trained to represent a step-dependent vector field that varies with a scalar variable --- interpreted as time / noise level in diffusion models or as time in flow matching formulations.
Because the network architecture itself is not inherently aware of this scalar, conditioning mechanisms are introduced to provide it with the corresponding value at each step.
Existing approaches can be grouped into three categories: (1) \textit{embedding-based conditioning}, which encodes the scalar variable into a learned embedding and injects it through dedicated conditioning modules, typically by adding it to the feature map~\cite{ho_NEURIPS2020_ddpm, song2021sde}, or by modulating normalization layers~\cite{dhariwal2021beat, nichol2021improved, peebles2023dit, esser2024stablediffusion3, chen2024gentron}; (2) \textit{input-level conditioning}~\cite{zhang2021dpir, lipman2022flowmatching1}, which augments inputs with constant-valued maps indicating the current step; and (3) \textit{external rescaling}~\cite{song2020improved_ncsnv2}, formulated for score-based diffusion models but not for flow matching, conditions the network by rescaling the predicted scores based on their known magnitude without modifying the network.

While each category provides a means for conditioning, they have distinct limitations. (1) \textit{embedding-based conditioning} has shown strong performance but relies on specialized architectural modules, such as normalization-based conditioning blocks, which constrain network design and require careful alignment of feature dimensionalities across normalization layers. In contrast, (2) \textit{input-level conditioning} and (3) \textit{external rescaling} are architecture-agnostic, but our experimental results indicate that they offer reduced expressiveness relative to embedding-based conditioning in denoising and unconditional image synthesis tasks. Moreover, (3) is less straightforward to extend to flow matching since it is designed specifically for score-based diffusion frameworks.
These trade-offs highlight the need for a conditioning strategy that is both architecture-independent and expressive across generative modeling frameworks.

\begin{figure*}[t]
\begin{center}
\includegraphics[width=1.\textwidth]{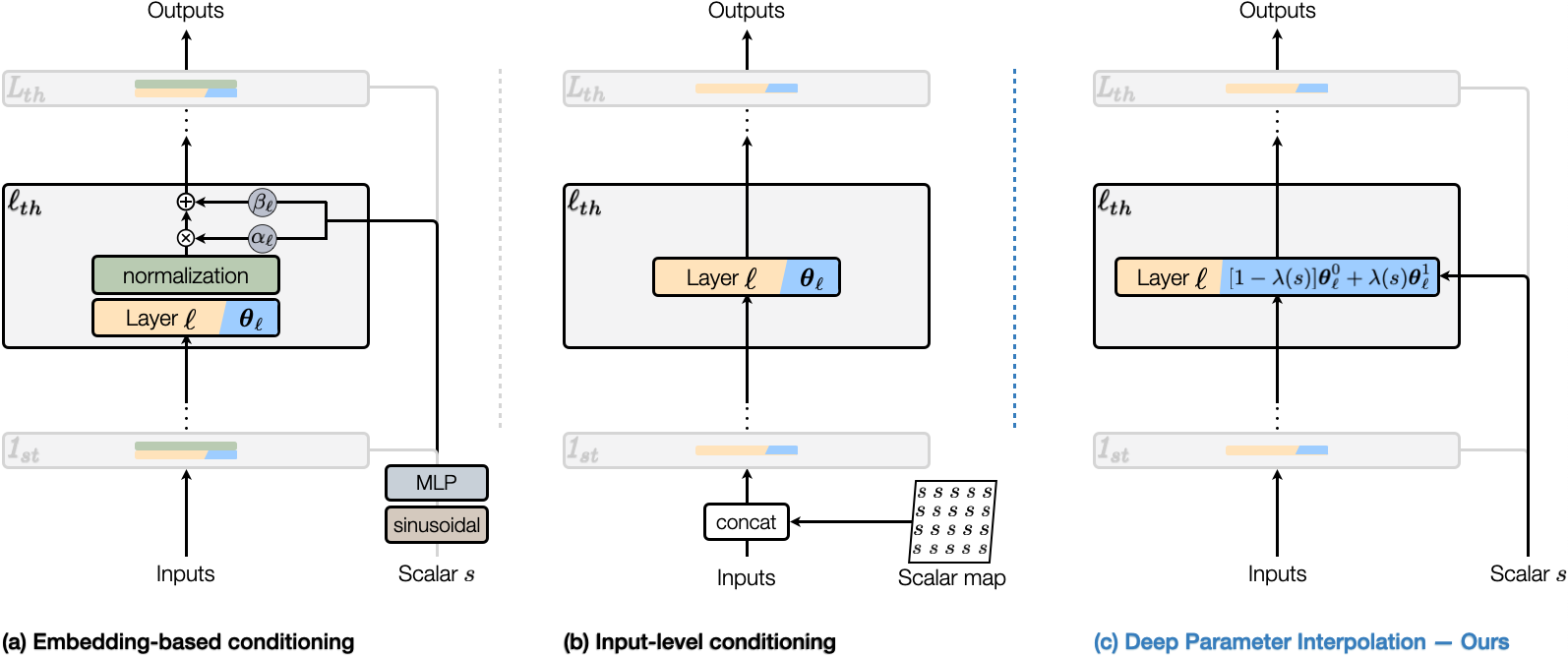}
\end{center}
\caption{Comparison of scalar conditioning mechanisms for diffusion and flow matching models.
Each method integrates a scalar variable $s$ (e.g., time or noise level) into the network, where learnable modules at layer $\ell$ are parameterized by $\vec{\theta}_{\ell}$.
\textbf{\textit{(a) Embedding-based conditioning}} injects a sinusoidal embedding of $s$ through an MLP to modulate features; effective but \textit{architecturally constrained}, as it requires conditioning modules (e.g., normalization layers).
\textbf{\textit{(b) Input-level conditioning}} concatenates a constant-valued scalar map with the input tensor; simple and architecture-agnostic but \textit{less expressive}.
\textbf{\textit{\textcolor{cvprblue}{(c) Deep parameter interpolation --- (ours)}}} maintains two learnable parameter sets, $\vec{\theta}^0$ and $\vec{\theta}^1$, and introduces scalar dependency at the parameter level by interpolating between them based on the scalar value $s$, where a learnable monotonic function $\lambda(s) \in [0,1]$ controls the interpolation to enable smooth adaptation across scalar values.}
\vspace{-.3cm}
\label{fig:illustration}
\end{figure*}

We introduce deep parameter interpolation (DPI), a general-purpose method for conditioning neural networks on scalar variables such as time or noise level.
Instead of injecting the scalar through input channels or requiring a specific architectural requirement to accept a scalar variable, we maintain two learnable parameter sets within a single network and dynamically interpolate between them according to the scalar value, introducing scalar awareness at the parameter level without altering the architecture.
Empirically, DPI achieves consistent improvements in denoising accuracy and sample quality over existing conditioning mechanisms, while achieving computational efficiency comparable to standard scalar conditioning techniques.

\section{Background}
\label{sec:background}

Modern generative modeling frameworks rely on parameterized stochastic or deterministic processes that progressively transform noise into structured data.
In this section, we introduce the score-based stochastic process and the flow-based deterministic process, along with the scalar conditioning methods used in both.

\subsection{Score-Based Diffusion Models}

Score-based diffusion models~\cite{song2019generative, ho_NEURIPS2020_ddpm, song2021sde, park2024randomwalks} learn the gradient of the log-density (score function) using neural networks. Tweedie’s formula~\cite{efron2011tweedie} connects the score function to the minimum mean squared error (MMSE) denoiser, enabling estimation of the score from noisy observations alone. Specifically, for noisy images \(\x_t = \x + \sigma_t \n\), where \(\x\) is clean image, \(\n \sim \mathcal{N}(\vec{0}, \vec{I})\), and \(\sigma_t\) denotes the noise level, the score can be approximated as
\[
\nabla \log p_{\sigma_t}(\x_t) \approx \frac{\mathsf{D}_\theta(\x_t\,;\,\sigma_t) - \x_t}{\sigma_t^2},
\]
where \(\mathsf{D}_\theta\) is a denoising network trained to minimize the mean squared error (MSE):
\[
\mathbb{E}_{\x, \n, t} \left[ \left\| \mathsf{D}_\theta(\x + \sigma_t \n \,;\, \sigma_t) - \x \right\|^2_2 \right].
\]
The scalar $\sigma_t$ (equivalently represented by the time variable $t$~\cite{park2024scorepnp}) thus acts as a conditioning signal that determines the current noise level during both training and sampling.

This formulation enables the denoiser to estimate the score function at various scalar noise levels, which makes it applicable for reverse-time sampling using stochastic processes~\cite{robbins1956empirical, miyasawa1961empirical, vincent2011connection}. The sampling process follows a stochastic random walk~\cite{ho_NEURIPS2020_ddpm, song2021sde, park2024randomwalks}, where each step involves moving the current state in the direction suggested by the estimated score function, combined with random noise.


\subsection{Flow Matching Generative Models}
\label{sec:flow_matching}

Flow matching models~\cite{lipman2022flowmatching1, albergo2023flowmatching2, liu2022flowmatching3} generate samples by learning continuous-time dynamics that transport samples from a
source distribution toward the data distribution.
Instead of estimating the score function, these models predict the velocity field $\vec{v}_\theta(\x_t\,;\,t)$ that describes how $\x_t$ evolves along a predefined probability path:
\[
\frac{d\x_t}{dt} = \vec{v}_\theta(\x_t\,;\,t).
\]

A common setup adopts an affine probability path~\cite{lipman2022flowmatching1}, where $\x_t$ is constructed as
\begin{equation}
\label{eq:flow_path}
\x_t = \beta_t \x_1 + \gamma_t \x_0,
\end{equation}
with $\x_1$ as a clean data sample, $\x_0 \sim \mathcal{N}(\vec{0}, \vec{I})$ as Gaussian noise, and $\beta_t$, $\gamma_t$ controlling the interpolation over time. Differentiating this path gives the target velocity:
\[
\vec{v}(\x_t\,;\,t) = \frac{d\beta_t}{dt} \x_1 + \frac{d\gamma_t}{dt} \x_0.
\]

The network $\vec{v}_\theta$ is trained to predict the target velocity by minimizing the MSE:
\[
\mathbb{E}_{\x_1, \x_0, t} \left[ \left\| \vec{v}_\theta(\x_t\,;\,t) - \vec{v}(\x_t) \right\|^2_2 \right].
\]
The scalar variable $t$ thus defines both the interpolation along the probability path and the conditioning context under which the velocity field is predicted.

This formulation enables the flow matching model to estimate the velocity field along the probability path, making it applicable for generative sampling. The sampling process evolves the state by following the learned velocity direction over continuous time, typically implemented using numerical ODE solvers.

\subsection{Scalar Conditioning in Generative Models}
\label{subsec:time_noise_conditioning}

Both score-based diffusion and flow matching models employ a single neural network that operates across a range of scalar conditions --- the noise level $\sigma_t \in [\sigma_1, \sigma_T]$ in diffusion models and the time step $t \in [0, 1]$ in flow matching. To make this possible, various conditioning mechanisms have been proposed to inform the network of the current step.

A common approach conditions the network by passing a sinusoidal embedding of the scalar variable \(s\) through a multi-layer perceptron (MLP) \(\varphi\) to produce modulation parameters
\[
    [a_\ell(s), b_\ell(s)] = \varphi_\ell(\mathsf{sinusoidal}(s)),
\]
where \(a_\ell(s)\) and \(b_\ell(s)\) are scale and shift coefficients applied at layer \(\ell\).
These parameters modulate the normalized activations as
\[
    h_{\ell+1} = a_\ell(s) \odot \mathsf{normalization}(h_\ell) + b_\ell(s),
\]
where \(h_\ell\) denotes the feature map at layer \(\ell\), and \(\odot\) represents element-wise multiplication.
While effective and popular, this design assumes the presence of normalization layers and ties the conditioning MLP to the channel dimensionality of each feature map, making it nontrivial to adapt the method to arbitrary network architectures.

Less constrained methods avoid architectural modifications by applying conditioning externally rather than through dedicated embedding modules. 
A simple approach applies conditioning at the input level by concatenating a constant-valued map that encodes the conditioning scalar \(s\) as the input tensor, $\x' = \mathsf{concatenate}\big(\x,\, s \cdot \mathbf{1}_{H \times W}\big)$,
where \(\x \in \mathbb{R}^{C \times H \times W}\) denotes the original input, and the constant map \(s \cdot \mathbf{1}_{H \times W}\) adds one channel encoding the scalar value. 
This formulation requires no architectural changes and is compatible with a wide range of backbones, though in practice we observe that its influence on feature representations tends to be relatively weak compared to embedding-based conditioning.

An alternative strategy, introduced in noise conditional score networks (NCSNv2)~\cite{song2020improved_ncsnv2}, removes explicit conditioning altogether by rescaling the predicted score \(\s_\theta(\x\,;\,\sigma)\) according to its magnitude with respect to the noise level \(\sigma\).
Empirically, the norm of the learned score satisfies $\|\s_\theta(\x\,;\,\sigma)\|_2 \propto 1/\sigma$,
which motivates the NCSNv2 parameterization
\[
\s_\theta(\x\,;\,\sigma) = \frac{1}{\sigma}\,\tilde{\s}_\theta(\x),
\]
where \(\tilde{\s}_\theta(\x)\) denotes an unconditional network output.
This formulation entirely removes architectural coupling; however, extending it to flow matching models remains nontrivial since they do not explicitly estimate scores, and our experiments further indicate that its conditioning capacity is limited.

\section{DPI: Deep Parameter Interpolation}
\label{sec:method}

\begin{figure*}[t]
    \centering
    \begin{minipage}[b]{0.45\textwidth}
        \centering
        \includegraphics[width=\textwidth]{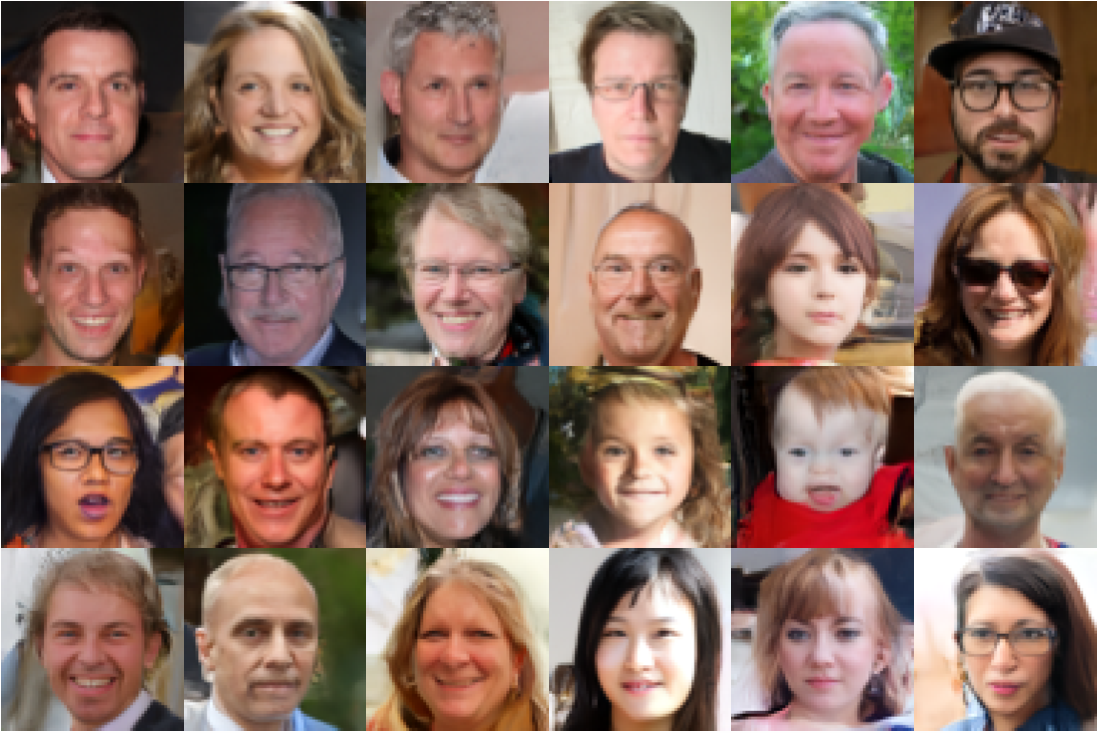}
        
        \small (a) DRUNet + DPI -- Diffusion SDE (FID: 12.23)
    \end{minipage}
    \hspace{0.02\textwidth}
    \begin{minipage}[b]{0.45\textwidth}
        \centering
        \includegraphics[width=\textwidth]{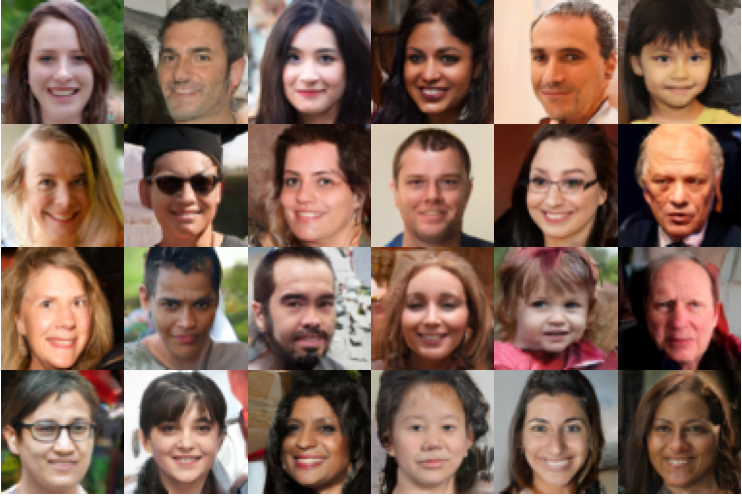}
        
        \small (b) DRUNet + DPI -- Flow ODE (FID: 12.57)
    \end{minipage}
    
    \vspace{0.3cm}
    
    \begin{minipage}[b]{0.45\textwidth}
        \centering
        \includegraphics[width=\textwidth]{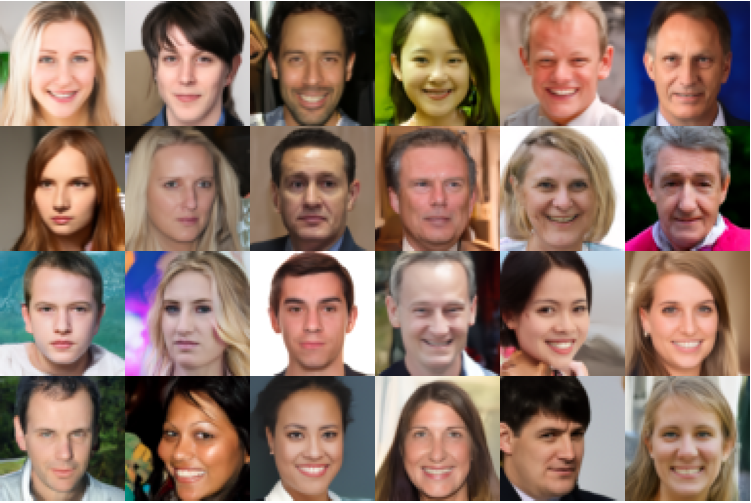}
        
        \small (c) ADM + DPI -- Diffusion SDE (FID: 8.32)
    \end{minipage}
    \hspace{0.02\textwidth}
    \begin{minipage}[b]{0.45\textwidth}
        \centering
        \includegraphics[width=\textwidth]{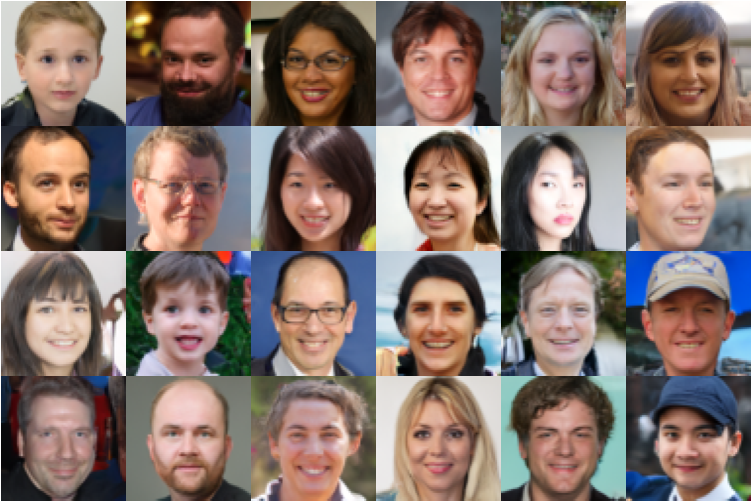}
        
        \small (d) ADM + DPI -- Flow ODE (FID: 7.96)
    \end{minipage}
    \caption{Examples of FFHQ $64 \times 64$ generated samples by DRUNet~\cite{zhang2021dpir} and ADM~\cite{dhariwal2021beat} using diffusion SDE and flow ODE solvers with our deep parameter interpolation (DPI). FID scores are reported in parentheses.}
    \label{fig:four_images_2x2}
\end{figure*}

We propose \textit{deep parameter interpolation (DPI)}, a simple yet effective strategy for conditioning neural networks on a scalar variable $s$ --- such as time or noise level --- directly at the parameter level. 
Instead of injecting scalar information through embeddings or input augmentation, our approach makes the model’s learnable parameters themselves vary smoothly with $s$.

To achieve this, DPI maintains two sets of learnable parameters for each learnable module (e.g., convolutional or linear layer) within a single network. It introduces scalar dependency at the parameter level by interpolating between these two parameter sets according to the scalar value.
We design the interpolation through a learnable monotonic function such that the interpolated parameter set is identical to the first parameter set at $s = s_{\min}$, identical to the second set at $s = s_{\max}$, and a linear combination of both for intermediate values $s \in (s_{\min}, s_{\max})$, enabling smooth network behavior across scalar values.
This parameter-level conditioning preserves the original architecture while adding negligible computational overhead, making it broadly applicable across diffusion, flow matching, and other scalar-conditioned frameworks with architectural flexibility.

Parameter interpolation in neural networks has been explored in other contexts,
e.g., in \cite{wang2019deepnetworkinterpolation},
which interpolates the parameters of separately trained networks for different objectives---one optimized for perceptual quality and another for distortion minimization.
Our approach is distinct in that the interpolation occurs within a single model during training, enabling scalar-conditioned adaptation without relying on multiple pre-trained networks.

\subsection{Interpolation of Learnable Modules for Scalar Conditioning}

Our key idea is that the vector field we aim to approximate varies smoothly with the scalar variable $s$.
To achieve this in the neural network, we enforce smoothness directly at the parameter level by letting parameters of each layer change smoothly with $s$ instead of keeping them fixed.

Concretely, let the base (i.e., not scalar-conditioned) network be defined as a function $f: \mathbb{R}^p \times \mathbb{R}^{m} \to \mathbb{R}^n ; (\bm{\theta}, \bm{x}) \mapsto f(\bm{\theta}, \bm{x})$,
where $\bm{\theta} \in \mathbb{R}^p$ are the network parameters (e.g., weights and biases) and $\bm{x}$ is the input (usually an image).
The DPI version of $f$ is a function 
$g: \mathbb{R}^{p} \times \mathbb{R}^{p} \times \mathbb{R}^{m} \times \mathbb{R} \to \mathbb{R}^n$
defined by
\begin{equation}
    g(\bm{\theta}^0, \bm{\theta}^1, \bm{x}, s) =f([1-\lambda(s)]\bm{\theta}^0 + \lambda(s)\bm{\theta}^1, \bm{x}),
\end{equation}
where $\lambda: \mathbb{R}\to \mathbb{R}$
is a monotonically increasing function such that $\lambda(s_\text{min})=0$ and
$\lambda(s_\text{max})=1$.

As a result of this design,
$g(\bm{\theta}^0, \bm{\theta}^1, \bm{x}, s_\text{min}) = f(\bm{\theta}^0, \bm{x})$
and
$g(\bm{\theta}^0, \bm{\theta}^1, \bm{x}, s_\text{max}) = f(\bm{\theta}^1, \bm{x})$;
%
 for intermediate values of $s$, the two parameter sets are smoothly blended by $\lambda(s)$.
This parameter level interpolation allows the network to adapt continuously across scalar values, introducing scalar awareness without modifying the underlying architecture (see Figure~\ref{fig:illustration}).

Although DPI maintains two learnable sets within a single network, its GPU memory and computational overhead remain comparable to a standard single-set network.
Interpolation involves only a lightweight element-wise combination of the two parameter sets, after which the model operates with a single set of interpolated weights.
As a result, the method achieves higher efficiency in GPU memory usage and comparable computation to embedding-based conditioning strategy such as MLP-based conditioning.
Further analysis is provided in Section~\ref{sec:efficiency}.

\subsection{Learnable Monotonic Interpolation Function}
\label{sec:learnable_interpolation_function}

The interpolation function $\lambda(s)$ determines how the network transitions as the scalar variable progresses from $s_{\min}$ to $s_{\max}$. 
It is defined to be monotonic and to satisfy
\[
\lambda(s_{\min}) = 0, \quad \lambda(s_{\max}) = 1,
\]
ensuring that the model initially relies on the first parameter set at $s_{\min}$ and gradually transitions to the $s_{\max}$ as \( s \) increases.

\begin{figure}[t]
\begin{center}
\includegraphics[width=.55\textwidth]{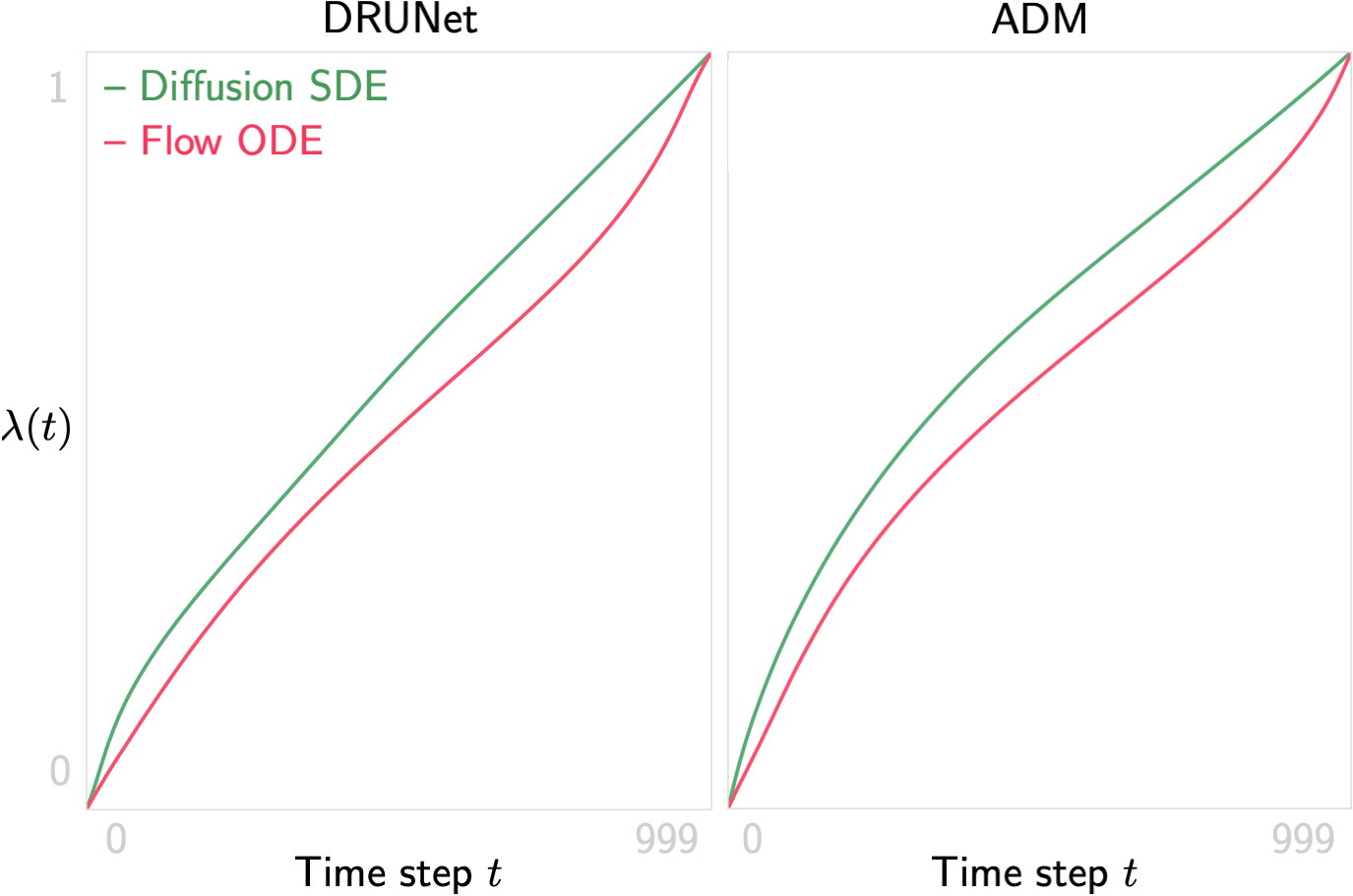}
\end{center}
\caption{Learned interpolation functions $\lambda$ for DRUNet and ADM under both diffusion and flow matching frameworks. The monotonic function $\lambda$ in Section~\ref{sec:learnable_interpolation_function} defines how the model interpolates two parameter sets within a single network as the scalar variable (time or noise level) progresses. Differences in $\lambda$ shapes indicate architecture- and framework-specific adaptation behavior.}
\label{fig:lambda_plot}
\end{figure}

We aim to make $\lambda(s)$ learnable while strictly enforcing its monotonicity across the scalar range. 
To achieve this, we design $\lambda(s)$ as a normalized cumulative distribution over a set of discrete scalar steps. 
This formulation allows flexible learning of the transition shape while guaranteeing that $\lambda(s)$ increases monotonically from $0$ to $1$.

Concretely, we introduce a learnable vector $\vec{\phi} \in \mathbb{R}^S$, where $S$ denotes the number of discrete scalar values (e.g., total number of timesteps or noise levels). 
We compute a softmax function over $\vec{\phi}$:
\[
p_i = \frac{\exp(\phi_i)}{\sum_{j=1}^{S} \exp(\phi_j)}, 
\]
where the softmax ensures that all $p_i$ are positive and sum to one.

We then define the interpolation function as the cumulative sum:
\[
\lambda_{\vec{\phi}}(s_i) = \sum_{j=1}^{i} p_j.
\]
This formulation guarantees $\lambda_{\vec{\phi}}(s_i) \in [0,1]$ and enforces strict monotonicity across the scalar range.

By learning this interpolation, the model continuously adapts its internal behavior across scalar values, effectively balancing denoising in diffusion models and velocity estimation in flow matching, without requiring explicit conditional inputs.

\section{Numerical Evaluations}
\label{sec:numericalevaluations}

\begin{table}[b]
    \setstretch{1.2} %
    \centering
    \footnotesize
    \caption{\small Architectural details of DRUNet~\cite{zhang2021dpir} and ADM~\cite{dhariwal2021beat} used in our generative modeling experiments.}
    \vspace{0.15cm}
    \renewcommand{\arraystretch}{1.0}
    \resizebox{0.4\textwidth}{!}{
    \begin{tabular}{@{}p{2.8cm}p{2.5cm}p{0.5cm}p{0.5cm}@{}p{0.2cm}@{}p{0.5cm}p{0.5cm}@{}}
    \toprule
    \noalign{\vskip -1.0ex}
    & \multicolumn{1}{c}{\textbf{DRUNet~\cite{zhang2021dpir}}}  
    & \multicolumn{1}{c}{\textbf{ADM~\cite{dhariwal2021beat}}}
    \\ \noalign{\vskip -.65ex}
    \cmidrule{1-3} \noalign{\vskip -.8ex}
    \textbf{$\#$ Parameters} & \multicolumn{1}{c}{63.9 M} & \multicolumn{1}{c}{60.9 M}  \\
    \textbf{$\#$ Residual blocks} & \multicolumn{1}{c}{8} & \multicolumn{1}{c}{1}  \\
    \textbf{Base channel width} & \multicolumn{1}{c}{64} & \multicolumn{1}{c}{128}      \\
    \textbf{$\#$ Attention heads} & \multicolumn{1}{c}{N/A} & \multicolumn{1}{c}{4}  \\
    \textbf{$\#$ Head channels} & \multicolumn{1}{c}{N/A} & \multicolumn{1}{c}{64}  \\
    \textbf{Attention resolutions} & \multicolumn{1}{c}{N/A} & \multicolumn{1}{c}{[16]}  \\
    \bottomrule
    \end{tabular}
    }
    \label{table:ablation_model_details}
\end{table}

We evaluate the proposed parameter interpolation in both diffusion and flow matching generative frameworks.
Our objectives are to (1) improve the denoising accuracy of diffusion models across a wide range of scalar conditions and (2) enhance unconditional image generation quality for both diffusion and flow matching models with comparable computational efficiency  to standard scalar conditioning techniques.
To this end, we conduct diffusion experiments that separately assess denoising and unconditional generation quality, and flow matching experiments focused on unconditional image generation.

\begin{figure*}[t]
\begin{center}
\includegraphics[width=.9\textwidth]{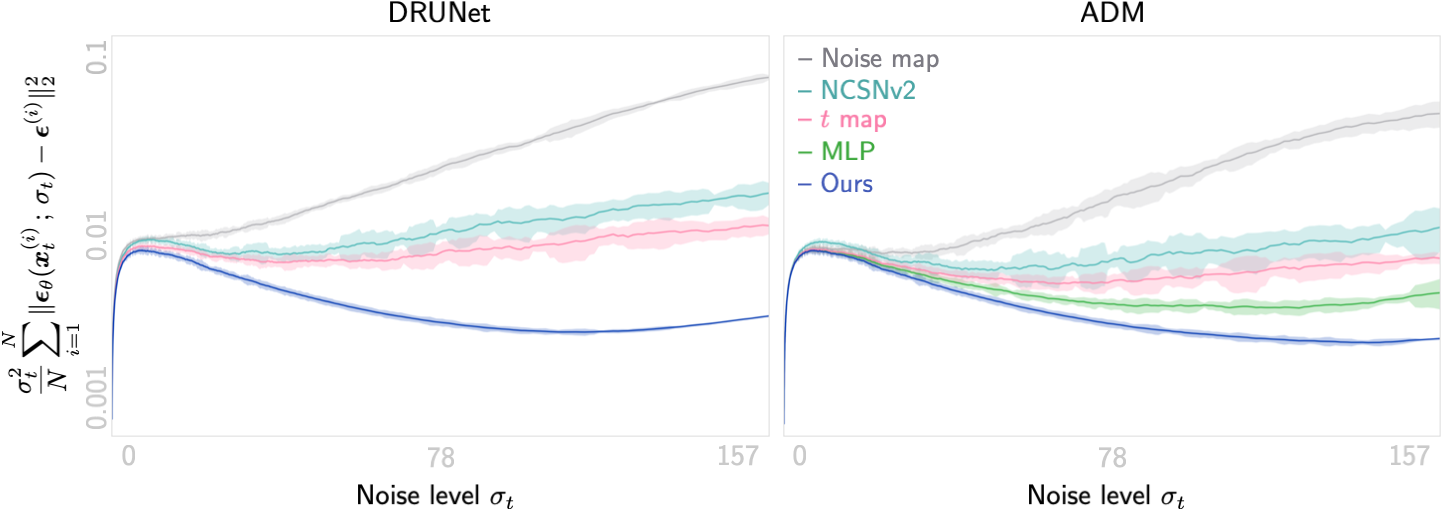}
\end{center}
\caption{Denoising performance across diffusion noise levels for different scalar conditioning methods using DRUNet and ADM. The plot shows the noise-scaled mean squared error (MSE), $\frac{\sigma^2_t}{N}\sum^N_{i=1} \| \vec{\epsilon}_{\theta}(\x_t^{(i)} ; \sigma_t) - \vec{\epsilon}^{(i)} \|^2_2$, where $\sigma_t^{2}$ denotes the noise variance at each diffusion step and $N$ the number of test samples. Our method consistently achieves lower error across a wide range of noise levels.}

\label{fig:denoising_plot}
\end{figure*}

\begin{table*}[t]
    \setstretch{1.35} %
    \centering
    \scriptsize
    \caption{\small Comparison of scalar conditioning methods for image denoising.
    We report the mean squared noise prediction error,
    $\frac{1}{N}\sum_{i=1}^{N}\|\vec{\epsilon}_{\theta}(\x_t^{(i)}\,;\sigma_t) - \vec{\epsilon}^{(i)}\|^2_2$, 
    where $N$ denotes the number of test samples. Results are shown at representative timesteps $t \in \{0, 333, 666, 999\}$ with corresponding noise levels 
    $\sigma_t \in \{0.01, 1.46, 9.49, 157\}$.
    The rightmost column shows the expected error over the full noise schedule. 
    \hlgreen{\textbf{Best}} and \hlblue{second-best} results are color-coded per noise level and per-architecture.}
    \vspace{-0.15cm}
    \renewcommand{\arraystretch}{1.2}
    \resizebox{\textwidth}{!}{
    \begin{tabular}{@{}>{\centering\arraybackslash}p{0.5cm}p{0.5cm}p{0.5cm}p{0.5cm}@{}p{0.2cm}@{}p{0.5cm}p{0.5cm}p{0.5cm}p{0.5cm}p{0.5cm}p{0.5cm}@{}}
    \toprule
    \noalign{\vskip -1.07ex}
    & &  \multicolumn{4}{c}{$\frac{1}{N}\sum_{i=1}^{N}\|\vec{\epsilon}_{\theta}(\x_t^{(i)}\,;\sigma_t) - \vec{\epsilon}^{(i)}\|^2_2$} & \\
    \cmidrule(lr){4-6}
    \noalign{\vskip -1.0ex}
    \multicolumn{1}{c}{\textbf{Sampling methods}} & \multicolumn{1}{c}{\textbf{Conditioning}} & \multicolumn{1}{c}{$t=0 \,(\sigma_t = 0.01)$} & \multicolumn{1}{c}{$t=333 \,(\sigma_t = 1.46)$} & \multicolumn{1}{c}{$t=666 \,(\sigma_t = 9.49)$} & \multicolumn{1}{c}{$t=999 \,(\sigma_t = 157)$} & \multicolumn{1}{c}{$\mathbb{E}\|\vec{\epsilon}_{\theta}(\x_t\,;\sigma_t) - \vec{\epsilon}\|^2_2$} \\ \noalign{\vskip -.65ex}
    \cmidrule{1-8} \noalign{\vskip -1.5ex}
    \multirow[c]{4}{*}{\textbf{DRUNet Diffusion}} & \multicolumn{1}{l}{NCSNv2} & \multicolumn{1}{c}{\hlblue{$5.282\mathrm{e}{-1}$}}  & \multicolumn{1}{c}{$1.331\mathrm{e}{-2}$}  & \multicolumn{1}{c}{$1.510\mathrm{e}{-3}$}  & \multicolumn{1}{c}{$1.123\mathrm{e}{-4}$} & \multicolumn{1}{c}{$2.038\mathrm{e}{-2}$}  \\ 
       & \multicolumn{1}{l}{$\sigma$ map} & \multicolumn{1}{c}{$5.562\mathrm{e}{-1}$}  & \multicolumn{1}{c}{$1.306\mathrm{e}{-2}$}  & \multicolumn{1}{c}{$1.522\mathrm{e}{-3}$}  & \multicolumn{1}{c}{$4.179\mathrm{e}{-4}$} & \multicolumn{1}{c}{$2.077\mathrm{e}{-2}$}  \\  
       & \multicolumn{1}{l}{$t$ map} & \multicolumn{1}{c}{$5.305\mathrm{e}{-1}$}  & \multicolumn{1}{c}{\hlblue{$1.249\mathrm{e}{-2}$}}  & \multicolumn{1}{c}{\hlblue{$1.410\mathrm{e}{-3}$}}  & \multicolumn{1}{c}{\hlblue{$7.854\mathrm{e}{-5}$}}  & \multicolumn{1}{c}{\hlblue{$1.987\mathrm{e}{-2}$}}  \\ 
       & \multicolumn{1}{l}{\textbf{Ours}} & \multicolumn{1}{c}{\hlgreen{$\bm{5.144\mathrm{e}{-1}}$}}  & \multicolumn{1}{c}{\hlgreen{$\bm{1.180\mathrm{e}{-2}}$}}  & \multicolumn{1}{c}{\hlgreen{$\bm{1.345\mathrm{e}{-3}}$}}  & \multicolumn{1}{c}{\hlgreen{$\bm{2.556\mathrm{e}{-5}}$}}  & \multicolumn{1}{c}{\hlgreen{$\bm{1.865\mathrm{e}{-2}}$}}  \\ 
       \cline{1-7}  \noalign{\vskip -.64ex}
       \multirow[c]{5}{*}{\textbf{ADM Diffusion}} & \multicolumn{1}{l}{NCSNv2} & \multicolumn{1}{c}{$4.914\mathrm{e}{-1}$}  & \multicolumn{1}{c}{$1.165\mathrm{e}{-2}$}  & \multicolumn{1}{c}{$1.453\mathrm{e}{-3}$}  & \multicolumn{1}{c}{$1.009\mathrm{e}{-4}$}  & \multicolumn{1}{c}{$1.795\mathrm{e}{-2}$}  \\ 
       & \multicolumn{1}{l}{$\sigma$ map} & \multicolumn{1}{c}{$4.970\mathrm{e}{-1}$}  & \multicolumn{1}{c}{$1.172\mathrm{e}{-2}$}  & \multicolumn{1}{c}{$1.390\mathrm{e}{-3}$}  & \multicolumn{1}{c}{$2.673\mathrm{e}{-4}$}  & \multicolumn{1}{c}{$1.819\mathrm{e}{-2}$}  \\  
       & \multicolumn{1}{l}{$t$ map} & \multicolumn{1}{c}{$4.922\mathrm{e}{-1}$}  & \multicolumn{1}{c}{\hlblue{$1.164\mathrm{e}{-2}$}}  & \multicolumn{1}{c}{$1.366\mathrm{e}{-3}$}  & \multicolumn{1}{c}{$5.346\mathrm{e}{-5}$}  & \multicolumn{1}{c}{\hlblue{$1.793\mathrm{e}{-2}$}}  \\  
       & \multicolumn{1}{l}{MLP} & \multicolumn{1}{c}{\hlblue{$4.853\mathrm{e}{-1}$}}  & \multicolumn{1}{c}{$1.171\mathrm{e}{-2}$}  & \multicolumn{1}{c}{\hlblue{$1.361\mathrm{e}{-3}$}}  & \multicolumn{1}{c}{\hlblue{$4.369\mathrm{e}{-5}$}}  & \multicolumn{1}{c}{$1.794\mathrm{e}{-2}$}  \\ 
       & \multicolumn{1}{l}{\textbf{Ours}} & \multicolumn{1}{c}{\hlgreen{$\bm{4.850\mathrm{e}{-1}}$}}  & \multicolumn{1}{c}{\hlgreen{$\bm{1.161\mathrm{e}{-2}}$}}  & \multicolumn{1}{c}{\hlgreen{$\bm{1.346\mathrm{e}{-3}}$}}  & \multicolumn{1}{c}{\hlgreen{$\bm{1.917\mathrm{e}{-5}}$}}  & \multicolumn{1}{c}{\hlgreen{$\bm{1.772\mathrm{e}{-2}}$}}   \\
       \cline{1-7}  \noalign{\vskip -.64ex}
    \bottomrule
    \end{tabular}
    }
    \label{table:denoising_epsilon}
\end{table*}

\subsection{Experimental Setup}
\label{sec:experimentsetup}

\textbf{Model Architectures.} 
We employ two representative architectures to demonstrate both compatibility and generality. 
First, the deep residual UNet (DRUNet)~\cite{zhang2021dpir}, a widely used image denoiser not originally designed for generative modeling, is included to demonstrate that our proposed conditioning enables such non-generative architectures to function effectively as generative models.
Second, the ablated diffusion model (ADM) U-Net~\cite{dhariwal2021beat} is selected for its open-source implementation of the sinusoidal embedding–based timestep conditioning that many later diffusion models build upon~\cite{nichol2021improved, peebles2023dit, chen2024gentron, esser2024stablediffusion3}. 
While the main objective of our experiments is to compare different scalar conditioning methods within each architecture, we additionally set the number of parameters of both architectures to be approximately equal---by adding more residual blocks to DRUNet---to enable a fair comparison across architectures as well.
The key architectural specifications for both models are summarized in Table~\ref{table:ablation_model_details}.

\textbf{Training Details.}
All models, including baselines, are trained from scratch on 69,000 FFHQ images ($64 \times 64$ RGB)~\cite{karras2019ffhq} for 500,000 iterations using a single NVIDIA RTX~A6000 GPU. 
We use the AdamW optimizer~\cite{loshchilov2019adamw} with learning rate $1 \times 10^{-5}$, batch size 256, dropout rate 0.1, and weight decay of 0.05. 
An exponential moving average (EMA) with decay rate 0.9999 is applied to stabilize optimization.
For our proposed parameter interpolation, which includes a learnable monotonic interpolation function $\lambda(s)$, we assign a separate learning rate of $1\times10^{-3}$ to the interpolation coefficients $\vec{\phi}$ (introduced in Section~\ref{sec:learnable_interpolation_function}).
All other hyperparameters are kept identical across models to ensure a fair comparison.

\textbf{Parameter Interpolation.} We apply the proposed parameter interpolation mechanism (Section \ref{sec:method}) to introduce scalar dependency at the parameter level. Specifically, we maintain two sets of learnable parameters and use a scalar-dependent, learnable interpolation function $\lambda(s)$ to linearly interpolate between them at each scalar step $s$, producing a single scalar-conditioned network configuration.

\begin{figure*}[t]
\begin{center}
\includegraphics[width=\textwidth]{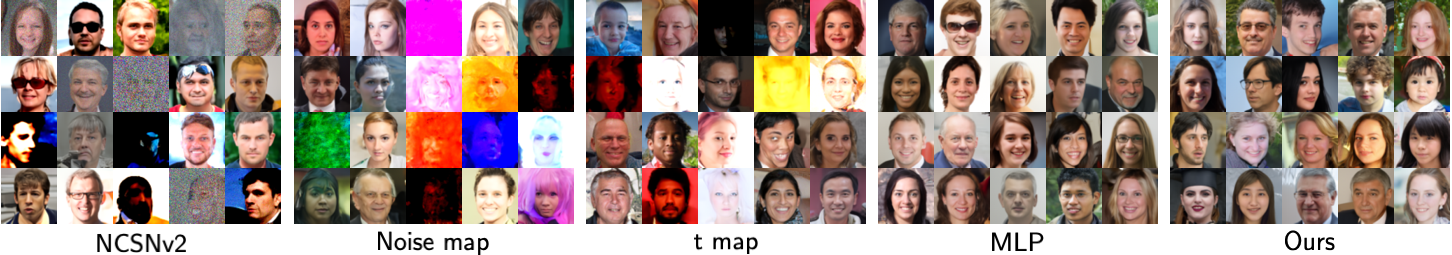}
\end{center}
\caption{Examples of FFHQ $64 \times 64$ generated samples by ADM~\cite{dhariwal2021beat} architecture using DDIM sampler~\cite{song2021ddim} with five different scalar conditioning methods.}
\label{fig:generatedsamples_by_methods}
\end{figure*}

\begin{table}[htbp]
    \setstretch{1.35}
    \centering
    \scriptsize
    \caption{\small Comparison of image generation quality across conditioning methods on both diffusion-based and flow matching-based sampling frameworks. 
    Our parameter interpolation achieves consistently superior performance in FID and sFID across all architectures, demonstrating effective and architecture-agnostic conditioning for generative modeling.
    \hlgreen{\textbf{Best}} and \hlblue{second-best} results are color-coded per sampling methods and per-architecture.}
    \vspace{0.15cm}
    \resizebox{0.5\textwidth}{!}{
    \begin{tabular}{@{}llcccc@{}}
    \toprule
 \textbf{\shortstack[l]{Sampling \\ methods}} & \textbf{Conditioning} & \textbf{FID}$\downarrow$ & \textbf{sFID}$\downarrow$ & \textbf{Precision}$\uparrow$ & \textbf{Recall}$\uparrow$ \\ 
    \cmidrule{1-6} 
         \multirow[c]{4}{*}{\shortstack[l]{\textbf{DRUNet} \\ \textbf{Diffusion}}}
        & NCSNv2 & $111.98$ & $68.13$ & $0.243$ & $0.265$ \\
         & $\sigma$ map & $137.77$ & $150.21$ & $0.089$ & $0.127$ \\
         & $t$ map & \hlblue{$26.23$} & \hlblue{$30.38$} & \hlblue{$0.457$} & \hlblue{$0.280$} \\
         & \textbf{Ours} & \hlgreen{$\bm{12.23}$} & \hlgreen{$\bm{17.11}$} & \hlgreen{$\bm{0.633}$} & \hlgreen{$\bm{0.307}$} \\ 
        \cdashline{1-6} 
         \multirow[c]{2}{*}{\shortstack[l]{\textbf{DRUNet} \\  \textbf{Flow} }}
         & $t$ map & \hlblue{$13.73$} & \hlblue{$16.60$} & \hlblue{$0.627$} & \hlblue{$0.302$} \\
         & \textbf{Ours} & \hlgreen{$\bm{12.57}$} & \hlgreen{$\bm{16.21}$} & \hlgreen{$\bm{0.635}$} & \hlgreen{$\bm{0.318}$} \\ 
    \cline{1-6}
    
         \multirow[c]{5}{*}{\shortstack[l]{\textbf{ADM} \\ \textbf{Diffusion} }}
        & NCSNv2 & $67.50$ & $47.35$ & $0.216$ & \hlgreen{$\bf{0.401}$} \\
         & $\sigma$ map & $96.71$ & $58.77$ & $0.283$ & $0.341$ \\
         & $t$ map & $13.14$ & \hlblue{$21.90$} & $0.581$ & $0.376$ \\
         & MLP & \hlblue{$10.14$} & $22.44$ & \hlgreen{$\bm{0.692}$} & $0.342$ \\
         & \textbf{Ours} & \hlgreen{$\bm{8.32}$} & \hlgreen{$\bm{17.45}$} & \hlblue{$0.667$} & \hlblue{$0.377$} \\ 
        \cdashline{1-6}
         \multirow[c]{3}{*}{\shortstack[l]{\textbf{ADM} \\ \textbf{Flow} }} 
         & $t$ map & \hlblue{$8.51$} & $16.41$ & \hlblue{$0.685$} & $0.373$ \\
         & MLP & $8.52$ & \hlblue{$16.25$} & $0.683$ & \hlblue{$0.381$} \\
         & \textbf{Ours} & \hlgreen{$\bm{7.96}$} & \hlgreen{$\bm{16.20}$} & \hlgreen{$\bf{0.687}$} & \hlgreen{$\bm{0.389}$} \\ 
    \bottomrule
    \end{tabular}
    }
    \label{table:fid}
\end{table}

For DRUNet, the interpolated parameters include all weights in 2D convolutional and transposed convolutional layers. For ADM, both convolutional weights and biases, as well as group-normalization parameters, are interpolated.
The interpolation function $\lambda(s)$ is defined by a learnable vector $\vec{\phi}$ of length $S = 1000$, matching the total timestep range used in both the diffusion and flow matching frameworks.
As shown in Figure~\ref{fig:lambda_plot}, the learned interpolation functions $\lambda(s)$ vary across architectures and generative frameworks, indicating that the model automatically adjusts its transition dynamics to match the scalar progression.

\subsection{Diffusion Framework Evaluations}
\label{subsec:diffusion_experiment}

In diffusion-based frameworks, we evaluate (1) denoising accuracy across noise levels and (2) unconditional image generation quality. 
These experiments validate whether our parameter interpolation can improve both step-wise noise handling and overall sampling performance.

\textbf{Training Setup.}
All diffusion models are trained under the variance-preserving (VP) formulation~\cite{ho_NEURIPS2020_ddpm, dhariwal2021beat}, where clean data $\x_0$ are progressively corrupted by Gaussian noise according to
\begin{equation}
\label{eq:vp_noise_addition}
    \x_t = \sqrt{\bar{\alpha}_t}\x_0 + \sqrt{1 - \bar{\alpha}_t}\vec{\epsilon},
\end{equation}
with $\vec{\epsilon} \sim \mathcal{N}(\vec{0}, \vec{I})$ representing the injected noise. 
Here, $\bar{\alpha}_t = \prod_{s=1}^{t} \alpha_s$ is the cumulative product of per-step noise coefficients $\alpha_s$, and $\alpha_t$ is scheduled linearly such that the variance increases from 0.0001 to 0.2 over training steps.
This schedule ensures that $\x_{t=0}$ corresponds to the data distribution, while $\x_{t=T}$ approaches a standard Gaussian.

Each diffusion model learns to predict the added noise $\vec{\epsilon}$ from the noisy observation $\x_t$, parameterized as $\vec{\epsilon}_\theta(\x_t\,;\,\sigma_t)$. 
The training objective minimizes the mean squared error between the predicted and true noise:
\[
\mathcal{L}_{\text{MSE}} = \mathbb{E}_{t, \vec{\epsilon}} \left[ \left\| \vec{\epsilon}_\theta(\x_t, \sigma_t) - \vec{\epsilon} \right\|^2_2 \right].
\]
All diffusion-based baselines follow this formulation, except NCSNv2~\cite{song2020improved_ncsnv2}, which directly estimates the score function $\nabla \log p_{\sigma_t}(\x_t)$ instead of the noise.

\textbf{Baseline Conditioning.} For ADM architecture, we test four noise conditioning schemes: NCSNv2~\cite{song2020improved_ncsnv2}, $\sigma$ map~\cite{zhang2021dpir}, $t$ map~\cite{lipman2022flowmatching1}, and MLP-based conditioning that modulates normalization layers~\cite{dhariwal2021beat}.
Among addition-based and normalization-based MLP conditioning variants, the normalization-based one is chosen, as it is adopted by many recent high-performing diffusion models~\cite{peebles2023dit, esser2024stablediffusion3, chen2024gentron}.
For DRUNet architecture, we test NCSNv2, $\sigma$ map, and $t$ map, omitting MLP-based conditioning due to the absence of normalization layers required for injecting sinusoidal embeddings of the scalar variable.

For NCSNv2 conditioning, we follow the original loss formulation in \cite{song2020improved_ncsnv2}, where the network directly predicts the score function, and apply a scaling term derived from the variance of the corresponding VP diffusion model.
The original NCSNv2, however, incorporates additional sampling refinements---an extra denoising step after annealed Langevin dynamics and a variance-exploding (VE) formulation with tuned step-size selection---which we omit to ensure a fair comparison under a consistent diffusion probabilistic sampling framework.

For the $\sigma$ map and $t$~map methods, the scalar variable is encoded as a spatially constant, 
single-channel map appended to the input. 
In the $\sigma$ map case, the conditioning value corresponds to the effective noise level associated with timestep $t$, defined as
\[
\sigma_t = \sqrt{\frac{1 - \bar{\alpha}_t}{\bar{\alpha}_t}},
\]
following~\cite{park2024scorepnp, park2024randomwalks}.
For the $t$~map method, the conditioning value is given by the normalized timestep $t/T$, 
which provides a scalar representation of the current diffusion step. 
Finally, MLP-based conditioning injects sinusoidal embeddings of $t$ into normalization layers via learned scale–shift parameters.

\textbf{Denoising Evaluation.} We assess denoising performance --- a fundamental indicator of diffusion model quality --- on 500 held-out images. 
Noisy image generation was repeated with 20 random seeds, and all conditioning methods share the same noisy image realizations to ensure fair comparison.
Figure~\ref{fig:denoising_plot} illustrates the mean squared error (MSE) of predicted noise $\vec{\epsilon}_{\theta}(\x_t)$ against the ground truth $\vec{\epsilon}$ across timesteps for DRUNet and ADM. 
Table~\ref{table:denoising_epsilon} summarizes representative results, showing that our proposed scalar conditioning consistently achieves the lowest prediction errors across both architectures.

\textbf{Sampling Evaluation.}
We generate 50,000 samples per model using 200 DDIM steps~\cite{song2021ddim} and evaluate FID~\cite{heusel2017gans_fid}, sFID~\cite{nash2021generating_sfid}, precision, and recall~\cite{kynkaanniemi2019improved_precisionrecall} using the public implementation from~\cite{dhariwal2021beat}, available at the following repository\footnote{\href{https://github.com/openai/guided-diffusion}{\textcolor{magenta}{https://github.com/openai/guided-diffusion}}}.
As shown in Figure~\ref{fig:generatedsamples_by_methods}, scalar conditioning methods with minimal architectural constraints (e.g., NCSNv2, $\sigma$ map, and $t$ map) can generate plausible faces but often produce artifacts or unstable exposure. 
In contrast, our method achieves stable and coherent visual quality without relying on dedicated scalar embedding modules, comparable to MLP-based conditioning that explicitly uses such embeddings.
Quantitative results in Table~\ref{table:fid} further confirm these improvements across all evaluation metrics.

\subsection{Flow Matching Framework Evaluations}
\label{subsec:flow_experiment}

\textbf{Training Setup.} Flow matching models are trained to predict velocity fields along the affine probability path~\cite{lipman2022flowmatching1},
\[
\x_t = \beta_t \x_1 + \gamma_t \x_0, \quad \text{with} \quad \beta_t = t, \, \gamma_t = 1-t,
\]
where $\x_1$ is a data sample and $\x_0 \sim \mathcal{N}(\vec{0}, \vec{I})$.
This formulation defines a linear interpolation between the data and Gaussian prior distributions, with the model learning the time-dependent velocity field that maps $\x_0$ to $\x_1$.

We use the same architectures, parameter settings, and conditioning schemes as in Section~\ref{subsec:diffusion_experiment}, except for NCSNv2~\cite{song2020improved_ncsnv2}, which is omitted because flow matching models directly predict velocity fields rather than score functions.

\textbf{Sampling Evaluation.} Consistent with diffusion framework evaluations, we generate 50,000 samples using the probability flow ODE with 200 steps. 
We evaluate the same perceptual metrics (FID, sFID, precision, and recall). 
As shown in Table~\ref{table:fid}, parameter interpolation again yields consistent quality improvements in FID and sFID across both ADM and DRUNet architectures, confirming that the proposed conditioning generalizes effectively beyond the diffusion formulation.

\begin{table}[htbp]
    \setstretch{1.35} %
    \centering
    \scriptsize
    \caption{\small Computational statistics of different conditioning methods.
    We report the number of parameters, peak GPU memory usage, and FLOPs for DRUNet and ADM under various scalar-conditioning schemes.
    Our parameter interpolation maintains efficiency comparable to baseline approaches despite doubling the parameter count.
    }
    \vspace{0.15cm}
    \renewcommand{\arraystretch}{1.35}
    \resizebox{0.5\textwidth}{!}{
    \begin{tabular}{@{}p{0.5cm}p{1.2cm}p{1.2cm}p{0.5cm}p{0.5cm}@{}}
    \toprule
    \noalign{\vskip -1.67ex}
    \multicolumn{1}{c}{\textbf{Models}}  & \multicolumn{1}{l}{\textbf{$t$ conditioning}} & \multicolumn{1}{c}{\textbf{Params (M)}} & \multicolumn{1}{c}{\textbf{Peak GPU (GB)}} & \multicolumn{1}{c}{\textbf{FLOPs (G)}}   \\ \noalign{\vskip -.65ex}
    \cmidrule{1-5} \noalign{\vskip -1.5ex}
    \multirow[c]{3}{*}{\textbf{DRUNet}}  & \multicolumn{1}{l}{NCSNv2} & \multicolumn{1}{c}{63.90} & \multicolumn{1}{c}{17.64} & \multicolumn{1}{c}{34.86} \\
       &  \multicolumn{1}{l}{$t$ map} & \multicolumn{1}{c}{63.90} & \multicolumn{1}{c}{17.65} & \multicolumn{1}{c}{34.86} \\
       &  \multicolumn{1}{l}{\textbf{Ours}} & \multicolumn{1}{c}{127.80} & \multicolumn{1}{c}{19.32} & \multicolumn{1}{c}{35.11} \\ \cline{1-5}  \noalign{\vskip -.64ex}
       \multirow[c]{4}{*}{\textbf{ADM}} & \multicolumn{1}{l}{NCSNv2} & \multicolumn{1}{c}{60.87}  & \multicolumn{1}{c}{37.24} & \multicolumn{1}{c}{49.54} \\
       &  \multicolumn{1}{l}{$t$ map} & \multicolumn{1}{c}{60.87}  & \multicolumn{1}{c}{37.25} & \multicolumn{1}{c}{49.54} \\ 
       &  \multicolumn{1}{l}{MLP} & \multicolumn{1}{c}{68.16}  & \multicolumn{1}{c}{41.82} & \multicolumn{1}{c}{49.56} \\ 
       &  \multicolumn{1}{l}{\textbf{Ours}} & \multicolumn{1}{c}{121.74}  & \multicolumn{1}{c}{38.84} & \multicolumn{1}{c}{49.77} \\  
       \cline{1-5}  \noalign{\vskip -.64ex}
    \bottomrule
    \end{tabular}
    }
    \label{table:computational_complexity}
\end{table}

\subsection{Computational Efficiency of Parameter Interpolation}
\label{sec:efficiency}

We compare computational efficiency in terms of floating-point operations (FLOPs) and peak GPU memory under identical batch and image settings (see Table~\ref{table:computational_complexity}).
Empirically, FLOPs increase by less than $0.72\%$, and peak GPU memory rises by approximately $5-9\%$, depending on the architecture. 
In contrast, MLP-based conditioning introduces additional per-layer transformations that noticeably raise GPU memory requirements. 
Overall, parameter interpolation achieves efficient conditioning with minimal overhead, preserving computation and memory efficiency while maintaining strong generative performance and broad architectural compatibility.

\section{Conclusion}
\label{sec:Conclusion}

We introduce \textit{deep parameter interpolation (DPI)}, a simple and general-purpose approach for conditioning neural networks on scalar variables such as time or noise level in diffusion and flow matching frameworks. 
Unlike existing conditioning mechanisms that require input modifications or specialized embedding layers, our method introduces scalar dependence directly at the parameter level by interpolating between two learnable parameter sets within a single network according to a learnable monotonic function. 
This design maintains the original architecture, making it broadly applicable across diverse generative frameworks.
Empirical evaluations on both diffusion and flow matching models demonstrate that DPI consistently improves denoising accuracy and sample quality across multiple architectures, including those not originally designed for generative modeling. 
Furthermore, the method achieves these gains with negligible computational overhead, offering a favorable balance between flexibility, efficiency, and performance.

\section{Acknowledgement}
\label{sec:acknowledgement}

This material is based upon work supported by the U.S. Department of Energy, Office of Science, Office of Advanced Scientific Computing Research under award number DE-SC0025589 and by the Laboratory Directed Research and Development program of Los Alamos National Laboratory under project number 20250637DI.
Additionally, this work was supported in part by the National Science Foundation under Grants No. 2504613 and
No. 2043134 (CAREER).


{
\small

\bibliographystyle{IEEEbib}
\bibliography{refs}
}

\newpage
\appendix

\section{Ablation: Learnable vs. Fixed Interpolation Functions}
\label{sec:learnable}

A central component of deep parameter interpolation (DPI) is the learnable monotonic interpolation function $\lambda(s)$, which determines how the model transitions between the two parameter sets across the scalar domain.
While Figure \ref{fig:lambda_plot} illustrates that $\lambda(s)$ adapts its shape depending on the architecture and generative framework, we additionally assess whether this learned flexibility is necessary or whether a fixed, non-learnable interpolation rule would be sufficient.

To evaluate this, we compare DPI with a fixed linear monotonic interpolation
\[
    \lambda(s) = s, \quad s \in [0, 1], 
\]
which removes all learnable parameters from the interpolation function while keeping all other aspects of DPI unchanged. 

We train DRUNet and ADM architectures under both diffusion and flow-matching settings using exactly the same configurations described in Section~\ref{sec:experimentsetup}. For each model, we compute the corresponding training objective on 500 held-out images across the whole scalar range (i.e., 1,000 scalar steps). At each scalar, we evaluate 20 noise realizations using shared random seeds to ensure fair comparison.

Table~\ref{table:ablation_objective} reports the averaged diffusion and flow objectives. In every setting, the learnable interpolation function achieves strictly lower error than the fixed linear rule, demonstrating that the ability to adapt $\lambda(s)$ is beneficial even though both versions interpolate between identical parameter endpoints.

\begin{table}[htbp]
    \setstretch{1.3}
    \centering
    \scriptsize
    \caption{\small Objective comparison between fixed linear and learnable interpolation functions. The learnable monotonic interpolation consistently reduces the diffusion and flow objectives compared to a non-learnable linear function. \hlgreen{\textbf{Best}} results are color-coded per sampling methods and per-architecture.}
    \vspace{0.15cm}

    \renewcommand{\arraystretch}{1.25}
    \resizebox{0.5\textwidth}{!}{
    \begin{tabular}{@{}p{.9cm}p{1.1cm}p{1.4cm}p{1.4cm}@{}}
        \toprule
        & & \multicolumn{1}{c}{\textbf{Diffusion Objective}} 
        & \multicolumn{1}{c}{\textbf{Flow Objective}} \\[-0.2em]
        \cmidrule(lr){3-3} \cmidrule(lr){4-4}
        \multicolumn{1}{c}{\textbf{Method}} 
        & \multicolumn{1}{c}{\textbf{$\lambda(s)$}} 
        & \multicolumn{1}{c}{{$\mathbb{E}\|\vec{\epsilon}_{\theta}(\x_t\,;\sigma_t) - \vec{\epsilon}\|^2_2$}}
        & \multicolumn{1}{c}{{$\mathbb{E}\|\vec{v}_{\theta}(\x_t\,; t) - \vec{v}\|^2_2$}} \\ 
        \midrule

        \textbf{DRUNet} & $s$ & \multicolumn{1}{c}{$1.870\mathrm{e}{-2}$} & \multicolumn{1}{c}{$1.250\mathrm{e}{-1}$} \\
        & Learnable & \multicolumn{1}{c}{\hlgreen{$\bm{1.865\mathrm{e}{-2}}$}}  & \multicolumn{1}{c}{\hlgreen{$\bm{1.209\mathrm{e}{-1}}$}} \\
        \midrule

        \textbf{ADM} & $s$ & \multicolumn{1}{c}{$1.775\mathrm{e}{-2}$} &  \multicolumn{1}{c}{$1.191\mathrm{e}{-1}$} \\
        & Learnable & \multicolumn{1}{c}{\hlgreen{$\bm{1.772\mathrm{e}{-2}}$}} & \multicolumn{1}{c}{\hlgreen{$\bm{1.086\mathrm{e}{-1}}$}} \\
        \midrule
    \end{tabular}
    }
    \label{table:ablation_objective}
\end{table}


\end{document}